\documentclass[aps,twocolumn,prl,
groupedaddress,nofootinbib,nobalancelastpage,nobibnotes]{revtex4}
\pdfoutput=1
\usepackage{amsmath,amsfonts,amssymb,mathrsfs,graphicx}
\usepackage[squaren]{SIunits}

\newcommand{\ie}{{\it i.e.}}

\newcommand{\eg}{{\it e.g.}}

\newcommand{\fig}{Fig.}

\newcommand{\Ref}{Ref.}
\newcommand{\Refs}{Refs.}

\newcommand{\figu}[1]{\fig~\ref{fig:#1}}

\newcommand{\bi}{\begin{itemize}}
\newcommand{\ei}{\end{itemize}}

\begin{document}

\title{Photohadronic Origin of the TeV-PeV Neutrinos Observed in IceCube}

\author{Walter Winter}
\affiliation{Institut f{\"u}r theoretische Physik und
  Astrophysik, Universit{\"a}t W{\"u}rzburg, Am Hubland, D-97074 W{\"u}rzburg, Germany}

\date{\today}

\begin{abstract}
We perform an unbiased search of the origin of the recently observed 28~events above $\sim 30 \, \mathrm{TeV}$ in the IceCube neutrino observatory, assuming that these are (apart from the atmospheric background) of astrophysical origin produced by photohadronic interactions. Instead of relying on the normalization of the neutrino flux, we demonstrate that spectral shape and flavor composition can be used to constrain or identify the source class. In order to quantify our observations, we use a model where the target photons are produced by the synchrotron emission of co-accelerated electrons, and we include magnetic field effects on the secondary muons, pions, and kaons. We find that the lack of observed events with energies much larger than PeV points towards sources with strong magnetic fields, which do not exhibit a direct correlation between highest cosmic ray and neutrino energies. While the simplest AGN models with efficient proton acceleration plausibly describe the current data at about the $3\sigma$ confidence level, we show that IceCube can rule out that the observed neutrinos stem from the sources of the ultra-high energy cosmic rays with a factor of ten  increased statistics at more than $5\sigma$ if the current observations are confirmed. A possible caveat are sources with strong magnetic fields and high Lorentz factors, such as magnetic energy dominated GRBs.

\end{abstract}

\maketitle

\section{Introduction}

High-energy astrophysical neutrinos probe the origin of the cosmic rays, since protons or cosmic ray nuclei interacting with ambient photons or non-relativistic matter will produce a high energy neutrino flux; see \Ref~\cite{Becker:2007sv} for a review. 
The recent discovery of two neutrino events at PeV energies by the IceCube neutrino telescope~\cite{Aartsen:2013bka} has therefore started the search for the sources of the cosmic rays from a new perspective, and the era of neutrino astronomy has begun. More recently, 26 additional events with deposited electromagnetic-equivalent energy between about 30 and 300~TeV from a contained vertex search based on data from  May 2010 to May 2012 have been presented~\cite{Nathan}. These events correspond to a 4.3$\sigma$ excess over the background of 10.6$^{+4.5}_{-3.9}$ from atmospheric neutrinos and muons. Since no clustering and no correlation with the galactic plane is significant from the data, it is plausible that the neutrinos (apart from the atmospheric background) come from extragalactic sources. 

In the meanwhile, the discussion of the origin of these events and the underlying spectral shape has gained momentum, see \Refs~\cite{Kistler:2013my,Cholis:2012kq,He:2013cqa,Laha:2013lka,Murase:2013rfa,Anchordoqui:2013lna,Fox:2013oza,Kalashev:2013vba,Stecker:2013fxa,He:2013zpa,Neronov:2013lza} for recent studies.  Cosmogenic neutrinos as a possible source are disfavored~\cite{Roulet:2012rv}, unless the maximal proton energy of the sources is below the threshold for the cosmic microwave background interactions, see \Ref~\cite{Kalashev:2013vba}. Many of the scenarios where the neutrinos are directly produced at the source assume a hadronuclear origin, which implies that the neutrino spectral shape follows the initial proton (or hadron) spectrum, and the maximal neutrino energy is limited by $E_{\nu, \mathrm{max}} \simeq 0.05 \, E_{p, \mathrm{max}}$ for protons. There are two puzzles in the current data: first of all, no neutrinos have been seen in the energy range between 250~TeV and 1~PeV, which may lead to the conclusion that the observed spectrum is a superposition of two groups of sources~\cite{He:2013zpa}. This observation may not yet be evident, since the deposited electromagnetic energy corresponds to a probability distribution of incident (true) neutrino energies; for instance, for a muon track, the true neutrino energy can be much higher. We therefore do not emphasize this aspect in this study.  Second,  no events above $~$PeV energies have been observed, which points towards a spectral break in the initial proton spectrum or a power law with a spectral index $\alpha \gtrsim 2.3$, see, \eg, \Ref~\cite{Anchordoqui:2013lna}. This seems to be in conflict with the typical assumptions for ultra-high energy cosmic ray (UHECR) accelerators, for which the primary spectrum has to extend to energies of about $10^{20} \, \mathrm{eV}$ and the injection index from Fermi shock acceleration is expected to be $2.0 \lesssim \alpha \lesssim 2.2$. 
However, it can be easily circumvented if a photohadronic origin of the neutrinos is assumed, as there is more freedom in the observed neutrino spectrum: first of all, the neutrino spectral shape will depend on the target photon spectral shape within the source as well, \ie, it does not have to follow the primary proton spectrum. Second, cooling effects of the secondary muons, pions, and kaons will affect the shape if the magnetic fields are strong enough, as, for instance, in microquasars~\cite{Reynoso:2008gs,Baerwald:2012yd}. These cooling effects will, most importantly, allow for a (moderate) decoupling of the maximal neutrino energy from the maximal proton energy, a fact which is highly relevant in exploring models in which the neutrinos come from the sources of the UHECRs.

 In this paper, we propose an unbiased search for the origin of the detected neutrinos from (presumably) astrophysical sources, based on spectral shape and flavor composition only, and assuming photohadronic production. We will especially address the question if the neutrinos can come from the sources of the UHECR, and if the simplest models for Active Galactic Nuclei (AGNs) can account for these neutrinos. We will, however, not describe the normalization of the observed neutrino spectrum, which typically has large astrophysical uncertainties, but instead just derive it as a free parameter. For the sake of simplicity, we assume that the observed neutrino spectrum comes from a single source class with similar parameters, which may be either of galactic origin or cosmologically distributed.

\section{Method and assumptions}

In order to reduce the complexity allowed by the freedom to choose the target photon spectrum, and to quantitatively support our (quite general) conclusions, we assume a pure proton spectrum at the source. We furthermore assume that the target photons are produced are produced by synchrotron emission of co-accelerated electrons, based on the model by {\em H{\"u}mmer et al.}~\cite{Hummer:2010ai}. 
For parameters corresponding to AGNs, the model gives neutrino spectral shapes similar to \Refs~\cite{Stecker:1991vm,Mucke:2000rn}. The protons and electrons are assumed to be injected from the acceleration zone into the radiation zone, where the interactions take place, with a power law spectrum with universal injection index $\alpha \simeq 2$. The maximal proton and electron energies are obtained by comparing the dominant energy loss timescale (adiabatic or synchrotron) to the acceleration rate $t^{-1}_{\mathrm{acc}}= \eta c^2 e B/E$, where $\eta \lesssim 1$ is the acceleration efficiency. In the radiation zone, secondary cooling effects are taken into account, as well as the relevant particle physics effects, which determine the shape and flavor composition. A more detailed discussion of the dominant effects on the neutrino spectra and flavor composition can be found in \Ref~\cite{Winter:2012xq}. As a consequence, the neutrino spectra then do not follow the initial proton spectrum anymore, neither in terms of spectral shape, nor in terms of maximal energy. Furthermore, the secondary cooling allows for flavor compositions different from $(\nu_e:\nu_\mu:\nu_\tau)=(1:2:0)$ (from the pion decay chain) at the source; see \Ref~\cite{Hummer:2010ai} for details. In order to fully describe the model including magnetic field effects and adiabatic cooling, the main parameters are $\alpha$ (injection index), $B$ (magnetic field), and $R$ (size of the region). Since the latter two parameters can be related to the Hillas plot~\cite{Hillas:1985is}, it is natural to present the results therein. The Hillas condition is automatically implied by  $t^{-1}_{\mathrm{acc}}=t^{-1}_{\mathrm{ad}}$ as a necessary condition for the maximal proton energy (which is, however, limited by synchrotron losses in parts of the parameter space). Flavor mixing is fully taken into account, using the best-fit parameters of \Ref~\cite{GonzalezGarcia:2012sz} with the first octant ($\theta_{23}$) solution. Note that we only discuss sources with low or moderate Doppler factors within this framework, as high Doppler factors complicate the interpretation in terms of the Hillas plot. 

The detector response to this model has been discussed in \Ref~\cite{Winter:2011jr}. Here we focus on the most recent data, using the exposures from \Ref~\cite{Anchordoqui:2013lna} (Fig~1) over 662 days of operation. Note that, for the sake of simplicity, we use the sky-averaged effective exposures here. Consider, for instance, $n$ sources with fluxes $J_\nu^i(E)$, and a declination-dependent exposure $\mathrm{Exp}_\nu(E,\delta)$, which includes the Earth attenuation of upgoing events.  Then the (total) number of events can be written as
\begin{equation}
N= \sum\limits_{i=1}^n \int dE \, \mathrm{Exp}_\nu(E,\delta^i) \, J_\nu^i(E) \, ,
\end{equation}
summed over the $n$ point sources. If the fluxes of the sources are similiar $J_\nu^i(E) \simeq J^{\mathrm{single}}_\nu(E)$, as it is assumed in this study, one can re-write this equation as
\begin{equation}
N= \int dE  \underbrace{\frac{1}{n} \sum\limits_{i=1}^n \mathrm{Exp}_\nu(E,\delta^i)}_{\mathrm{Exp}^{\mathrm{eff}}_\nu(E)} \, \underbrace{n \, J^{\mathrm{single}}_\nu(E)}_{J_\nu(E)} \, ,
\end{equation}
and use an appropriately averaged exposure.
If moreover the sources are uniformly distributed over the sky, as it is typically assumed for extragalactic sources, one can also derive the averaged exposure as the solid angle average
\begin{equation} 
\mathrm{Exp}^{\mathrm{eff}}_\nu(E) = \frac{1}{2} \int d\cos \delta \, \,  \mathrm{Exp}_\nu(E,\delta) \, ,
\end{equation}
which we imply in this study.

The atmospheric neutrino flux is taken as the one measured by IceCube~\cite{Aartsen:2012uu}. For the binning, we choose four bins in the {\em reconstructed} neutrino energy: 30 to 200~TeV, 200 TeV to 1 PeV, 1 to 3 PeV, and 3 PeV to 100 PeV to have a meaningful number of events per bin. For the sake of simplicity, we assume that the electromagnetic equivalent energy is roughly 25\% of the incident neutrino energy for a muon track, and 75\% for a cascade~\cite{Laha:2013lka}; cascades from neutral current interactions are assumed to be suppressed by the cross sections and their reconstruction at lower energies. We have tested two different assumptions: either cascades and tracks are added into one event sample, or separate samples for cascades and tracks have been chosen (default), without qualitative differences -- see discussion below.  The event rates are computed from the spectral shapes of all three flavors at Earth convoluted with the flavor-dependent exposures. The (Poissonian) $\chi^2$ using the observed 28~events and the fit rates is obtained by minimizing the fit normalization as a free parameter. As a consequence, we have reproduced the results from \Ref~\cite{Anchordoqui:2013lna}, such as that spectral indices harder than 2.3 for a single power law are ruled out. For the atmospheric neutrino background we obtain 8.8 events, in sufficient consistency with the total background obtained in the IceCube analysis. In the absence of further information, we assume the reconstructed event topology distribution (cascades versus tracks) follows the one produced by atmospheric neutrinos, although a part of the actual background comes from atmospheric muons sneaking in through the dust layer~\cite{Claudio}, and a part of the atmospheric neutrino background is reduced by the veto of atmospheric muons produced in the same air shower.

\section{Results}

\begin{figure}[t!]
\begin{center}
\includegraphics[width=0.9\columnwidth]{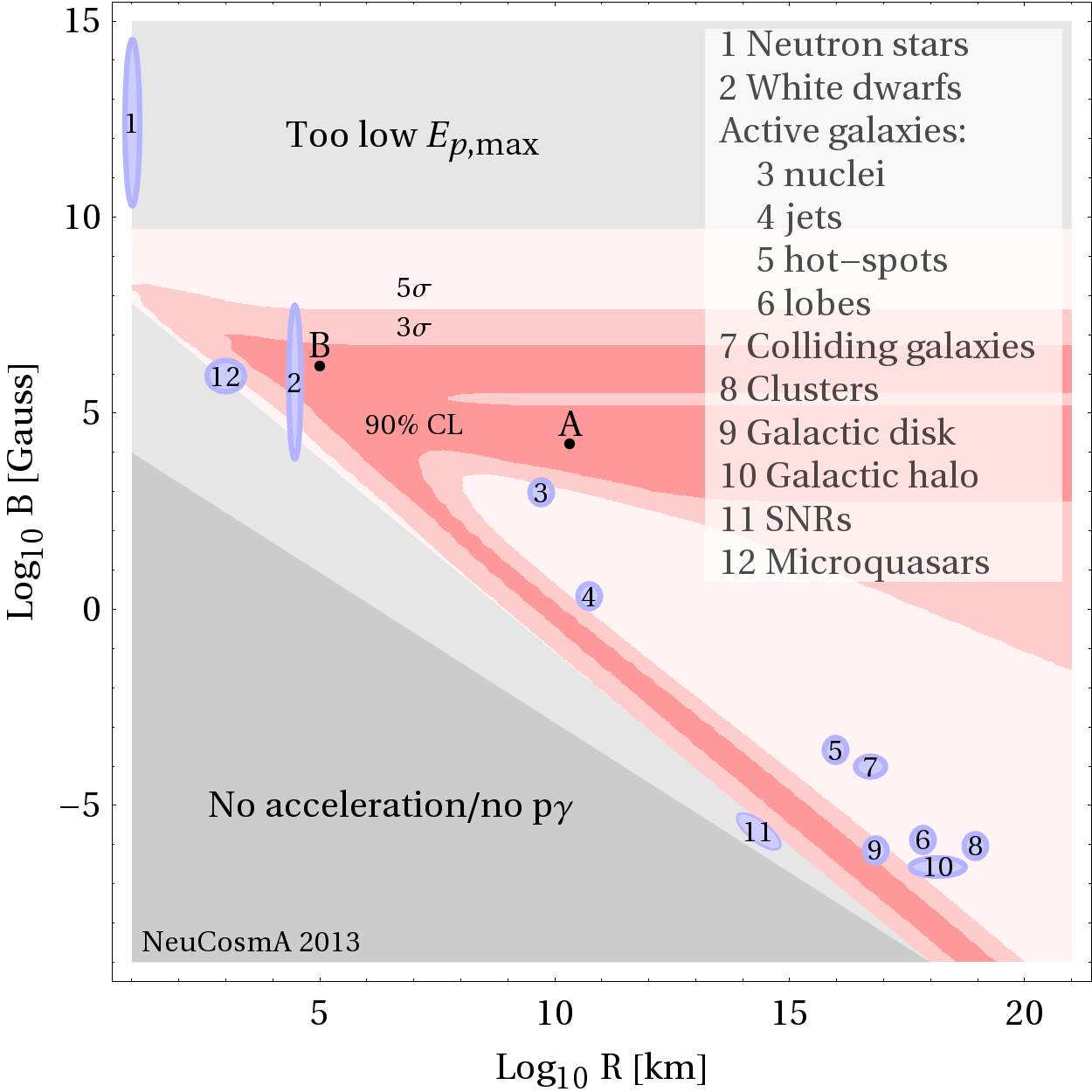}
\vspace*{-0.5cm}
\end{center}
\caption{\label{fig:hillas} Allowed regions (90\% CL, 3$\sigma$, 5$\sigma$; 2 d.o.f.) in the Hillas plane from the recently observed 28 IceCube events, assuming a cosmological distribution of identical sources following the star formation rate,  an injection index for protons and electrons $\alpha \simeq 2$, and a (moderate) acceleration efficiency $\eta=0.1$. The test points~1 to~11 have been adopted from M. Boratav for illustration (see \eg\ \Ref~\cite{Ostrowski:2001ej}), test point~12 from \Ref~\cite{Baerwald:2012yd}, and points~A and~B mark the best fits.  }
\end{figure}

\begin{figure*}[t!]
\begin{center}
\begin{tabular}{ccc}
\includegraphics[width=0.3\textwidth]{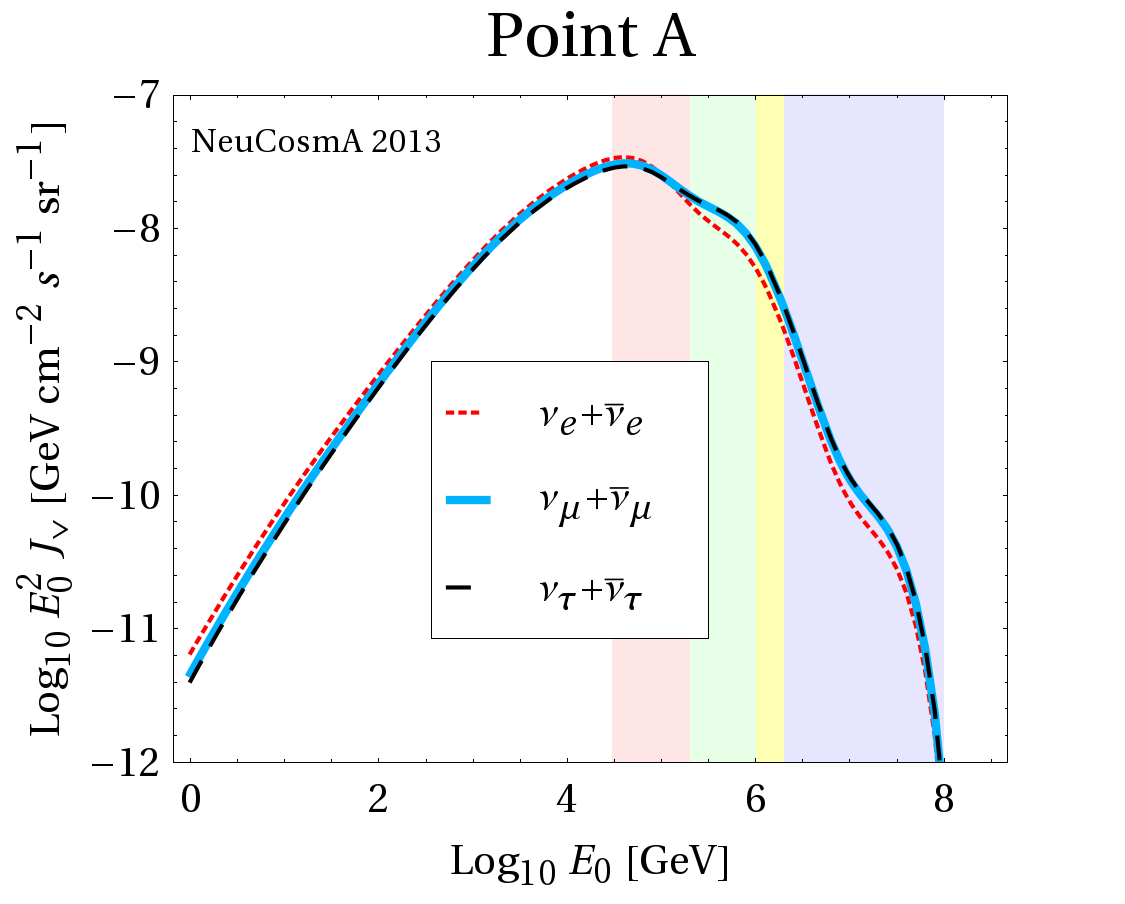} & \includegraphics[width=0.3\textwidth]{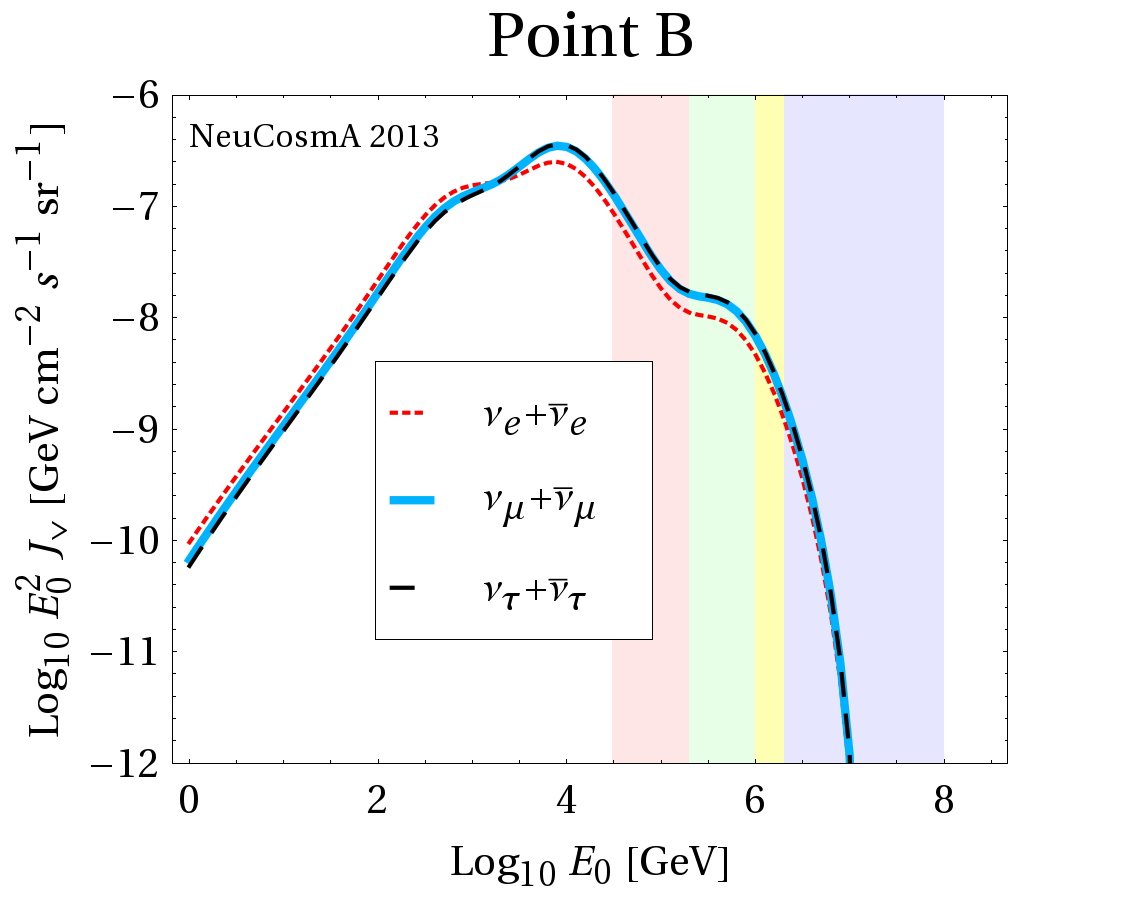} &
\includegraphics[width=0.3\textwidth]{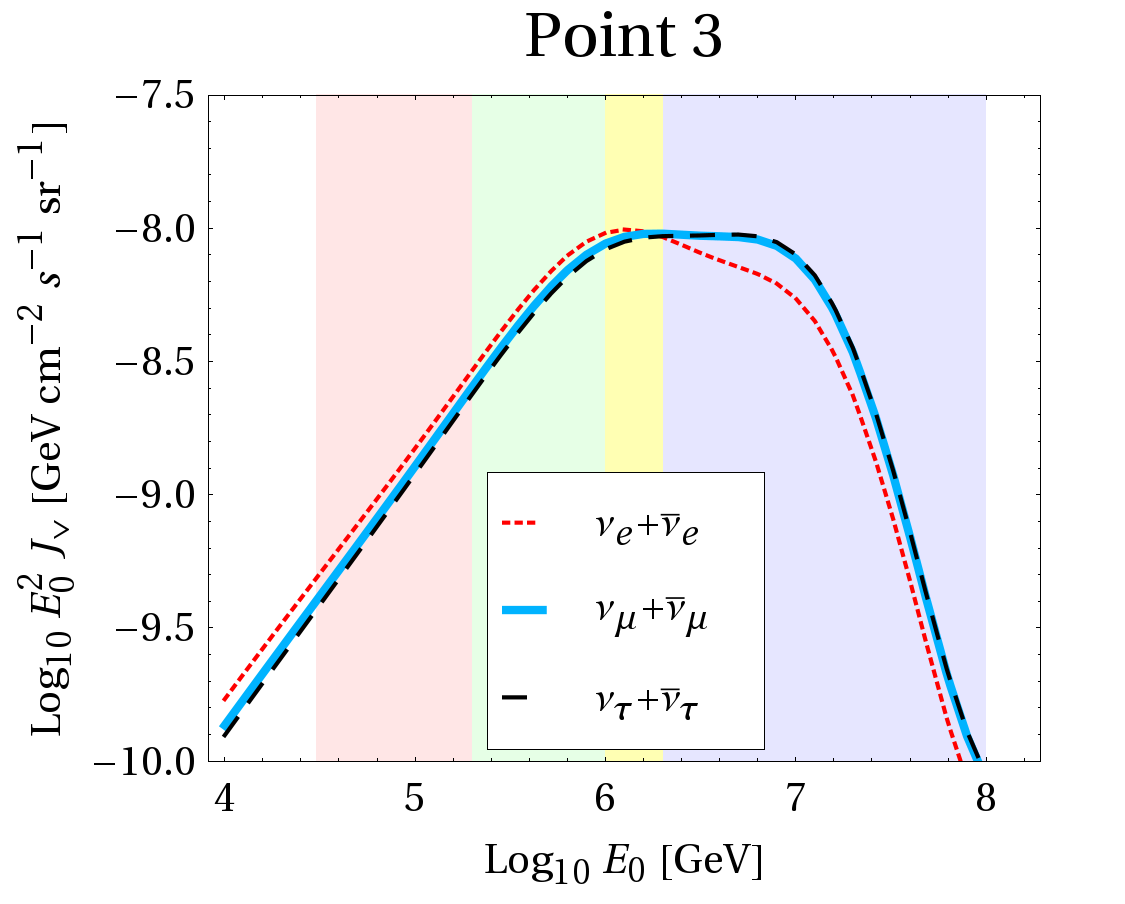}  \\
\includegraphics[width=0.3\textwidth]{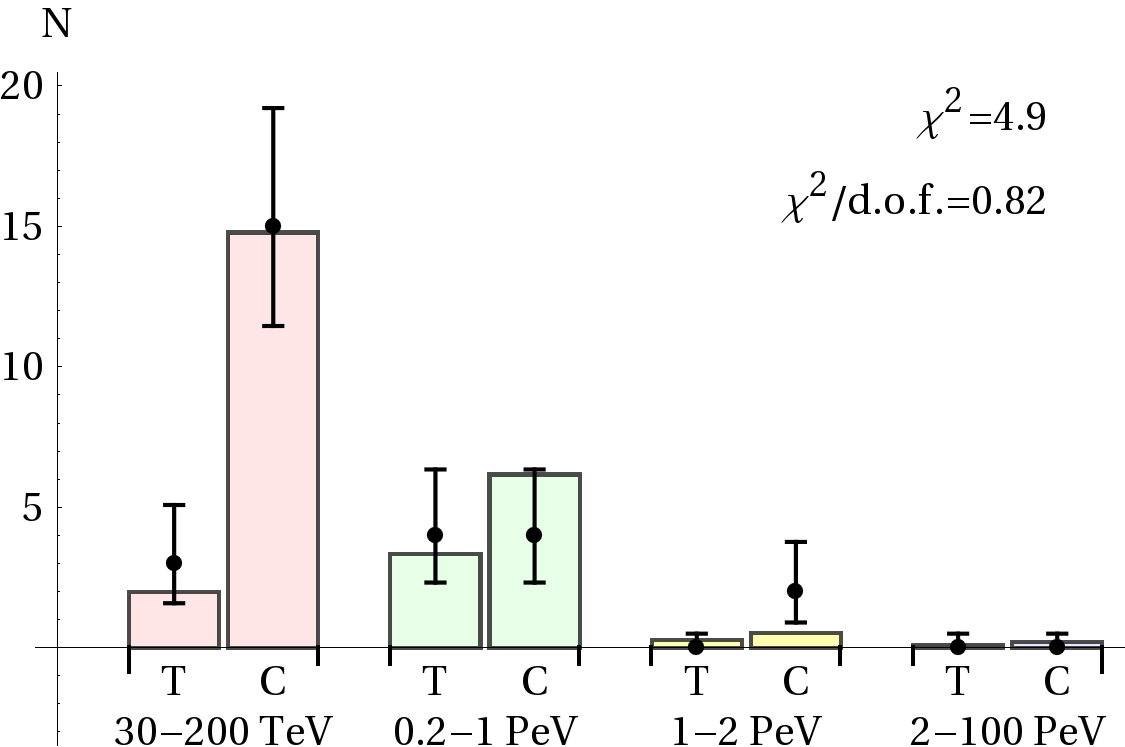}  &
\includegraphics[width=0.3\textwidth]{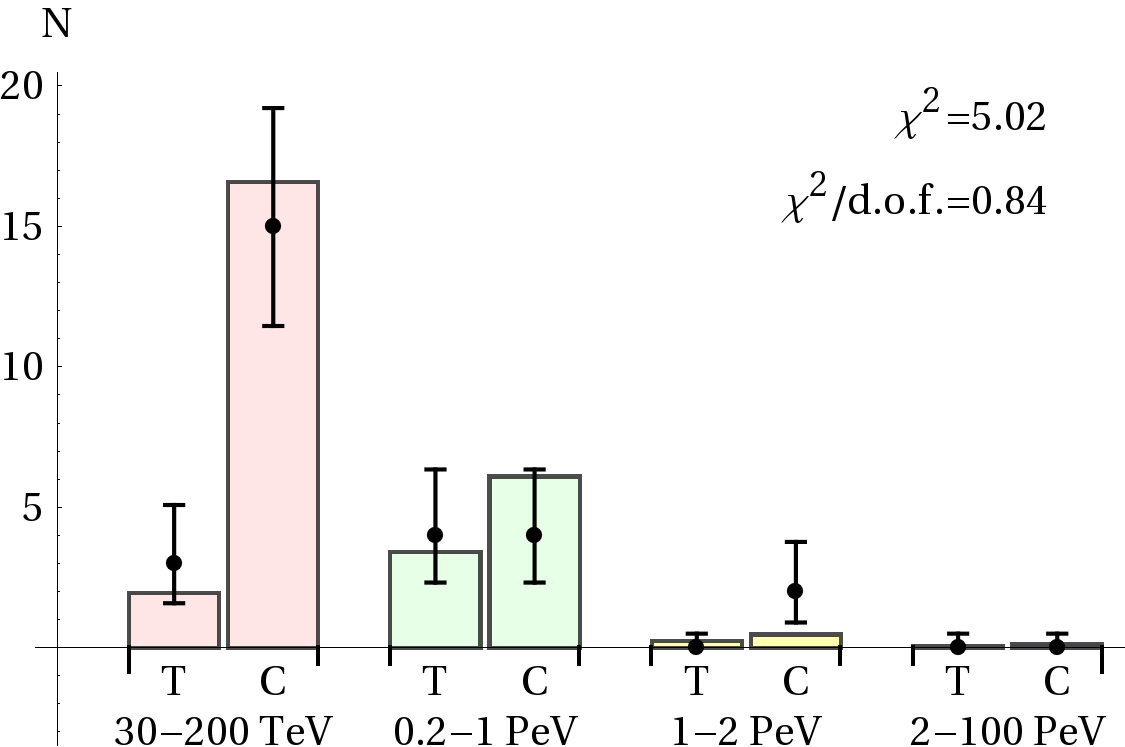}  &
\includegraphics[width=0.3\textwidth]{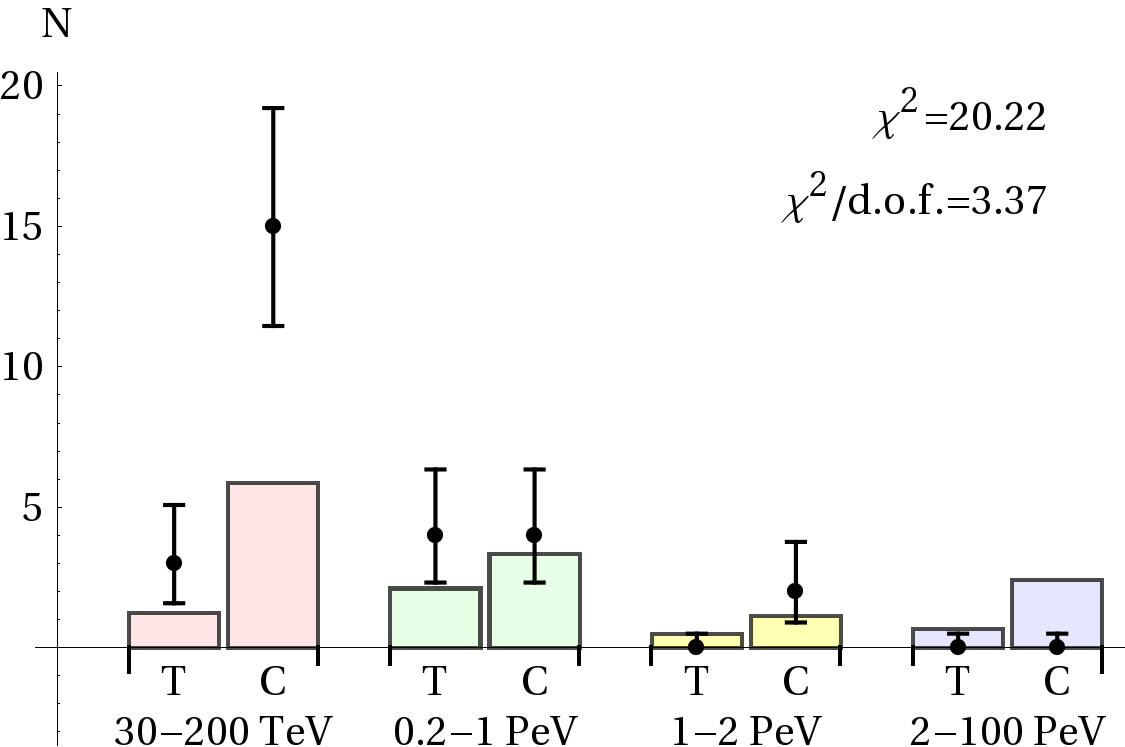}  \\
\end{tabular}
\end{center}
\caption{\label{fig:bestfit} Upper row: best-fit neutrino spectra for points~A,~B, and~3 in \figu{hillas}  as a function of the reconstructed neutrino energy at the observer $E_0$ (assuming star formation rate evolution). Lower row: observed (dots) and fitted (bars) event rates in the different bins with Poissonian error bars for muon tracks (T) and cascades (C). The energy ranges for the bins are also shaded by corresponding colors in the first row. }
\end{figure*}

A parameter space fit as a function of $R$ and $B$ for the 28~events can found in \figu{hillas}, where the sources are assumed to be cosmologically distributed following the star formation rate~\cite{Hopkins:2006bw} and atmospheric backgrounds are fully taken into account. The best-fit points are marked by~A and~B, and a number of additional test points (adopted from M. Boratav, see \eg\ \Ref~\cite{Ostrowski:2001ej}),  are shown. Note that the exact values for these sources are not to be taken too seriously, since large parameter variations are possible. 
In the figure, the 90\% CL, 3$\sigma$, and 5$\sigma$ contours are shown for 2 d.o.f., as well as the regions incompatible with the photoproduction threshold (``No acceleration/no $p\gamma$'') and  with the proton energy required to produce the neutrinos in the first bin (``Too low $E_{p,\mathrm{max}}$'').
The $\chi^2$ per degree of freedom at the minima is roughly one, which means that model provides a good fit. 
For $B \lesssim 10^3 \, \mathrm{G}$, where magnetic field effects on the secondaries are small, the fit is determined by the neutrinos following the maximal proton energy, whereas for $B \gtrsim 10^3 \, \mathrm{G}$, magnetic field effects on the secondaries are more important; see \Refs~\cite{Hummer:2010ai,Winter:2011jr}. One can read off from the figure that at $5 \sigma$, none of the allowed points can be excluded with current data, whereas parameter sets corresponding to supernova remnants (SNRs) and objects beyond galactic scales do not match the required energy range at $3\sigma$.

The best-fit points in \figu{hillas} can be found in regions with large magnetic fields, whereas the test points~3 and~4 (AGN nuclei and jets, respectively) are on the boundary of the $3\sigma$ contour.  In order to illustrate what happens there, we show in \figu{bestfit} the spectra (upper row) and rates (lower row) for best-fit points~A and~B, as well as test point~3.  The bin ranges are marked in the upper row as correspondingly shaded regions. For a discussion of the magnetic field and flavor effects leading to these spectra, see \Ref~\cite{Hummer:2010ai}.
First of all, we recover from points~A and~B that a steep cutoff is needed to match the non-observation of events beyond a few PeV. In cases~A and~B, the spectral peak is within and below the observed energy range, respectively; in case~B, it is hidden by the atmospheric background. The event rates fit the observation extremely well, apart from a small mismatch in the 1-2 PeV bin by observing two cascades in that energy range and no tracks. The magnetic field effects can be clearly seen as ``wiggles'' in the upper panels, which also leads to a change of the flavor composition as a function of energy. The inferred flux normalizations and shapes are compatible with the predictions of models for specific objects in the literature, such as for AGNs (Point~3, \Refs~\cite{Stecker:1991vm,Stecker:2005hn}) or Pulsar Wind Bubbles (Point~B, \Ref~\cite{Granot:2002qz}). We have tested the impact of the flavor measurement (muon tracks versus cascades), which is in slight tension with the spectral shape information. For instance, using the flavor information only, the region $B \gtrsim 10^5 \, \mathrm{G}$ could be excluded with a factor of 100 higher statistics. For realistic exposures, however, we find that the flavor information is not competitive at all, which is due to the small impact of the initial flavor composition after flavor mixing, and the limited energy range accessible in the current analysis. This conclusion may change in the presence of new physics in the neutrino propagation (such as neutrino decay or quantum decoherence), which can significantly alter the flavor composition; see \eg\ \Ref~\cite{Mehta:2011qb} in the context of the model used here.

The spectral shape and normalization of the AGN spectrum for test point~3 (upper right panel in \figu{bestfit}) are consistent with the prediction of typical AGN models (here without luminosity distribution function), such as the Stecker et al. model~\cite{Stecker:1991vm,Stecker:2005hn,Stecker:2013fxa}. From the lower right panel, one can easily read off where the statistical tension for these models comes from: there are too few events predicted at low energies and too many events at high energies compared to the observation. While the low energy part may be argued away by the superposition of two different groups of sources~\cite{He:2013zpa}, the high energy part cannot be easily circumvented. Thus, either sufficiently many events at even higher energies are going to be found soon, or this part of the parameter space using this simple model will be ruled out. Note that these observations cannot be easily applied to AGNs in general. For instance, there are indications from particle-in-cell simulations that the maximal energies could be overestimated~\cite{Sironi:2010rb,Sironi:2013ri}, which would improve the neutrino fit, but increase the tension with description of the UHECR observations. In addition, superimposing different populations of AGNs with different characteristics may help to fit all data. Nevertheless, this example illustrates that neutrino spectral data alone can be a powerful model discriminator.

\begin{figure}[t!]
\begin{center}
\includegraphics[width=0.9\columnwidth]{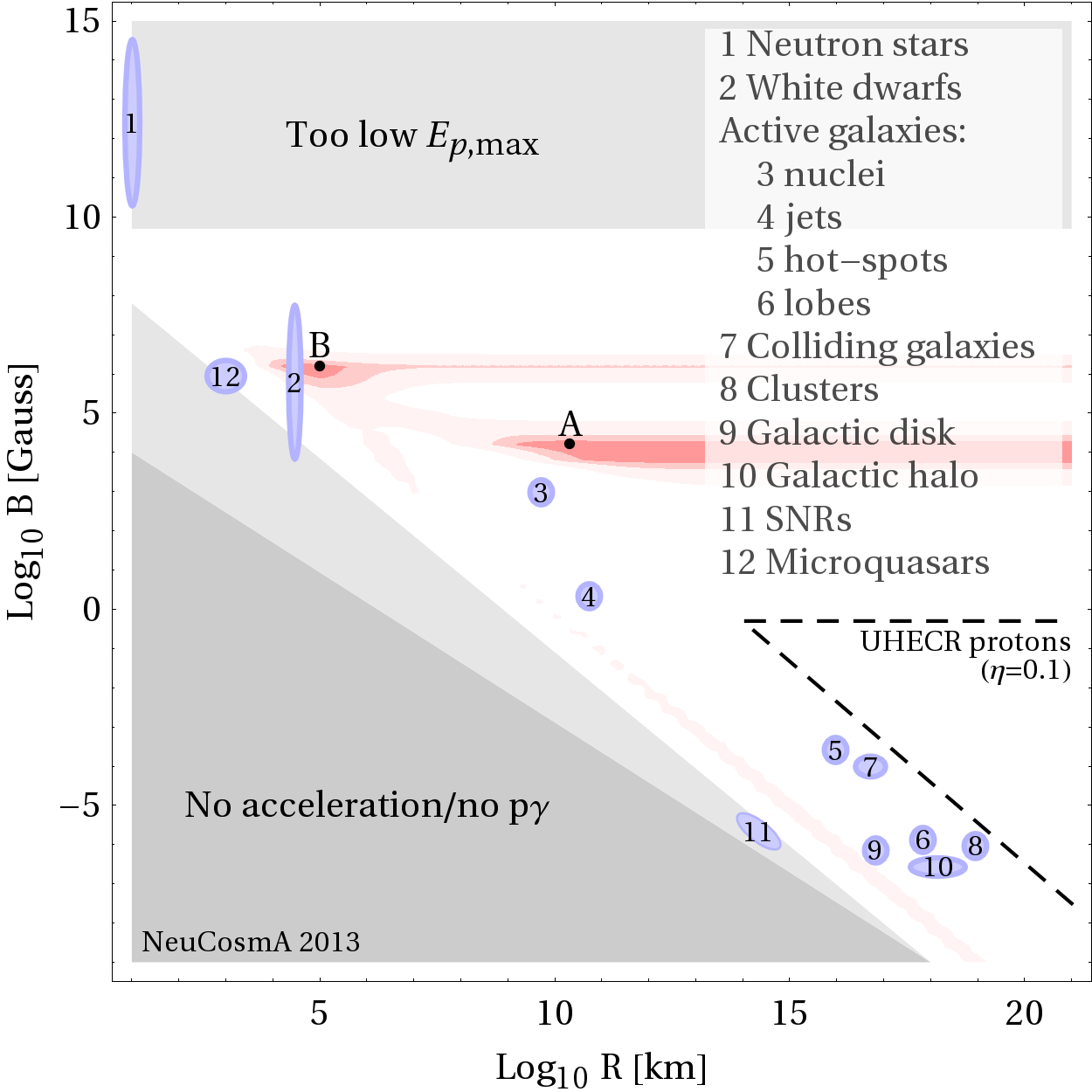}
\vspace*{-0.5cm}
\end{center}
\caption{\label{fig:hillasexp}  Same as \figu{hillas} with an increase of statistics by a factor of ten. Here it is assumed that the current observation will be confirmed, \ie, is not just a statistical fluctuation.
The region ``UHECR protons'' marks where UHECR protons with $E_p \gtrsim 10^{20} \, \mathrm{eV}$ are expected to come from in the model for the chosen acceleration efficiency.}
\end{figure}

In order to quantify this statement, we show in \figu{hillasexp} the expected fit for a factor of ten increased statistics, as it may come from running the full scale experiment over a sufficiently long time. In this figure, it is assumed that the current statistical distribution of events is representative, \ie, future events will follow this distribution. One can clearly see that in that case points~A and~B are remaining,  but that almost the full parameter space with $B \lesssim 10^{3.5} \, \mathrm{G}$ can be ruled out with an impressing precision -- including the simplest AGN models.
 In particular, note that the UHECR in this model come from the dashed region. The lower edge of this region just follows the Hillas criterium for cosmic ray protons (Larmor radius smaller than acceleration region), the upper edge corresponds to the necessary condition that the acceleration timescale must be smaller than the synchrotron loss timescale. The region somewhat depends on acceleration mechanism, acceleration efficiency, and UHECR composition. Nevertheless, it is clear that if the current data are confirmed, IceCube will rule out at a high confidence level that the observed neutrinos come from the sources of the UHECR, because the best-fit and UHECR regions are completely disjunct. On the other hand, sources with relatively strong magnetic fields remain viable, where the secondary cooling breaks the connection between ultra-high primary and neutrino energy. In the discussed model, however, these sources will not be capable to accelerate protons to the required energies because of the synchrotron loss limitation. A possible way out may be sources with strong magnetic fields and high Lorentz factors, such as Gamma-Ray Bursts (GRBs), which can allow for strong secondary cooling and high proton energies at the same time; see \eg\ spectrum in \Ref~\cite{Hummer:2011ms}, which resembles the ones in the left and middle upper panels in \figu{bestfit}, apart from peaking at higher energies. Since $E_{p,\mathrm{max}} \propto (B')^{-1/2}$ and the synchrotron cooling break of the secondaries scales as $E_{\mathrm{br}} \propto (B')^ {-1}$, where  $B'$ is the magnetic field in shock rest frame, large values $B' \gtrsim 10 \,\mathrm{MG}$ can be used to suppress disproportionally the secondary cooling break to the PeV range while maintaining large maximal proton energies $E_{p,\mathrm{max}} \gtrsim 10^{10} \, \mathrm{GeV}$ through large enough Lorentz boosts $\Gamma \sim 300$; see \eg\ \Ref~\cite{Baerwald:2011ee} for detailed formulae. However, note that such high magnetic fields are significantly higher (by about a factor of 50) than what is typically assumed within the internal shock model for conventional long duration GRBs. A possible corresponding GRB scenario involves ultra-long GRBs, see \Ref~\cite{Murase:2013ffa}, for which, however, the maximal proton energy is limited by $pp$ and $p\gamma$ cooling. 

We have also studied how the conclusions depend on the model parameters. We have tested different values of the injection index $\alpha$ and different source evolution functions, such as the one for AGNs in \Ref~\cite{Hasinger:2005sb} and the no evolution case, as expected if the neutrinos come from our galaxy. We have also tested the impact of a different event topology composition of the atmospheric muon background, and the impact of a different binning including an intermediate bin with no events.
Although the final result slightly depends on the assumptions, there are no qualitative changes of our conclusions. In particular, in all cases, the overall best-fit lies close to test point~B.  

\section{Summary and conclusions}

Assuming a photohadronic origin of the TeV-PeV astrophysical neutrinos recently observed by the IceCube collaboration, we have studied what can be learned from spectral shape and flavor composition of the neutrinos. 
In order to quantitatively support our conclusions, we have assumed that protons at the source are injected  with a non-thermal power law spectrum with injection index two. The target photons have been assumed to come from synchrotron emission from co-accelerated electrons, as it is often assumed for AGNs. Most importantly, magnetic field effects on the secondary muons, pions, and kaons have been fully taken into account, which determine the neutrino spectra for strong magnetic fields, as well as flavor mixing, high energy multi-pion production processes, and the helicity dependence of the muon decays, which can also affect the spectral shape and flavor composition. 

We conclude that current data prefer source classes with strong magnetic fields $B \gtrsim 10^{3.5} \, \mathrm{G}$, while the simplest AGN models are still allowed at the $3\sigma$ confidence level. This information can be inferred from spectral shape and flavor composition only. The spectral
shape clearly provides the main constraint, while flavor (in the sense
of the muon track to cascade ratio) has little discrimination power yet -- at least  in the absence of new physics effects altering the neutrino propagation. 

With increased statistics, IceCube will allow an unprecedented view on the parameter space, which potentially includes source classes (or parameter sets) which are unobservable in photons. For example, 
if the current observations are confirmed with about a factor of ten increased statistics, the parameters corresponding to AGNs in the discussed model, as well as to other sources classes with magnetic fields $B \lesssim 10^{3.5} \, \mathrm{G}$ will be ruled out if the cosmic ray acceleration is efficient. 

More generally and quite independent of the model assumptions, the current non-observation of events above a few PeV either points towards  strong magnetic field effects on the secondaries (muons, pions, kaons) in order to break the direct connection between maximal neutrino energy and maximal cosmic ray energy ($E_{\nu, \mathrm{max}} \simeq 0.05 \, E_{p, \mathrm{max}}$ for protons), or to significantly lower acceleration efficienices of the cosmic rays than anticipated earlier. In either case, IceCube will then be able to exclude that the observed neutrinos come from the sources of the UHECR, since either the required magnetic fields would limit the maximal cosmic ray energy by synchrotron losses, or lower acceleration efficiencies could not describe the UHECR production itself. A possible caveat may be sources with strong magnetic fields and high Lorentz factors at the same time, such as GRBs with substantial magnetic field energies.

\subsection*{Acknowledgments}

I would like to thank Philipp Baerwald, Mauricio Bustamante, Nayantara Gupta, Svenja H{\"u}mmer, Michele Maltoni, and Kohta Murase for useful discussions and comments.
This work has been supported by DFG grants WI 2639/3-1 and WI 2639/4-1, the FP7 Invisibles network (Marie Curie
Actions, PITN-GA-2011-289442), and the ``Helmholtz Alliance for Astroparticle Physics HAP'', funded by the Initiative and Networking fund of the Helmholtz association.

\vspace*{0.5cm}


\end{document}